\documentclass[allclo]{FBSart}
\usepackage{amsfonts}
\usepackage{amssymb}
\usepackage{epsfig}

\newcommand{\case}[2]{\mbox{$\frac{#1}{#2}$}}
\title{Manifestation of three-body forces
in three-body Bethe--Salpeter and light-front equations}
\author{V.A.\ Karmanov$^{a}$ and P.\ Maris$^{b}$}
 \institute{$^a$Lebedev Physical Institute, Leninsky
Prospekt 53, 119991 Moscow, Russia\\ $^b$Department of Physics and
Astronomy, Iowa State University, Ames, Iowa 50011, USA}

\runningauthor{V.A.\ Karmanov and P.\ Maris}
\runningtitle{Manifestation of three-body forces...}
\begin{document}

\maketitle
\begin{abstract}
Bethe--Salpeter and light-front bound state equations for three scalar
particles interacting by scalar exchange-bosons are solved in ladder
truncation.  In contrast to two-body systems, the three-body binding
energies obtained in these two approaches differ significantly from
each other: the ladder kernel in light-front dynamics underbinds by
approximately a factor of two compared to the ladder Bethe--Salpeter
equation.  By taking into account three-body forces in the light-front
approach, generated by two exchange-bosons in flight, we find that
most of this difference disappears; for small exchange masses, the
obtained binding energies coincide with each other.
\end{abstract}
%
%
%
%
%
%
%
%
%


\section{Introduction
\label{intro}}
Relativistic two-body systems are often treated using the ladder
truncation in the Bethe--Salpeter (BS) framework
\cite{Maris:2003vk} or in light-front dynamics (LFD)
\cite{cdkm,bpp}. These two approaches give results very close to
each other for the binding energy \cite{mc_2000} and also for
electromagnetic form factors \cite{kcm06} in the spinless case.
Binding energies of a two-fermion system were calculated both in
the BS approach in \cite{dorkin} and in LFD \cite{mck03} and again
the results turned out to be rather close to each other.

In these cases, the relativistic calculations using either the BS
approach or LFD lead to binding energies that are quite different from
the non-relativistic binding energy, obtained by solving the
Schr\"odinger equation.  The relativistic effects are more important
for larger binding energy.  In addition, they are also important for
small binding energy, if the mass of the exchange particle is large
enough, that is, of the order of the constituent mass \cite{mc_2000}.

Here we extend the same methods, the BS approach and LFD, to scalar
three-body bound states, using a one-boson exchange kernel.
Eventually, the goal is to extend these methods to bound states of
fermions: e.g. using a meson-exchange model in the case of 3-nucleon
bound states (triton, ${}^3$He), and baryons as bound states of three
(non-perturbatively dressed) quarks interacting via gluons.
For simplicity however, and as a first step, we restrict ourselves
here to spinless systems.  Preliminary results for the three-body BS
equation with a one-boson exchange kernel have been presented in
\cite{Pieter}.  As far as we know, the LFD three-body bound state
equations with one-boson exchange kernel has never been solved for
such systems.  Previously, the LFD equation was solved in
\cite{tobias,ck_03} for zero-range interaction.  Its solution was also
found in the relativistic quantum-mechanical approach for scattering
states \cite{polyzou}, with a phenomenological mass operator.

It is important to note that the actual ladder kernels in the BS
approach and in LFD are not identical, nor are they given by the same
graphs.  In the BS ladder kernel, there is no notion of
time-ordering in the diagrams.  On the other hand, the kernel for the
LFD equation is given by the time-ordered graphs in the light-front
(LF) time.

\begin{figure}[tb]
\begin{center}
\includegraphics[width=8cm]{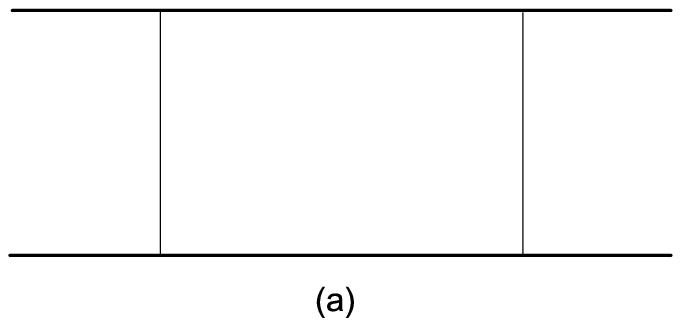}
\includegraphics[width=8cm]{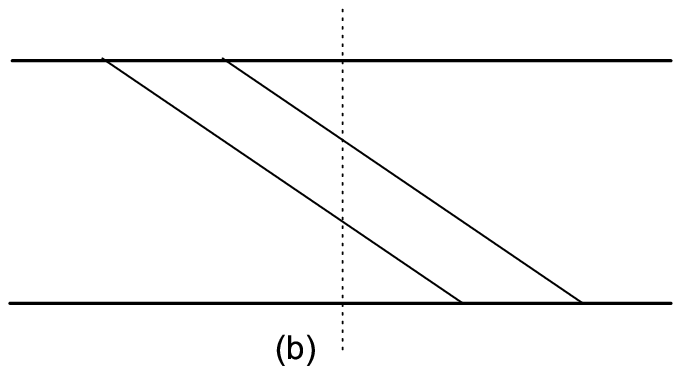}
\caption{(a) Top: Feynman ladder graph with two exchanges.  (b)
Bottom: One of six time-ordered graphs, generated by the ladder
Feynman graph (a). \label{feyn2}} \end{center}
\end{figure}
Thus, the second iteration in the two-body ladder BS equation (the
second-order Feynman box graph, Fig.~\ref{feyn2}(a)), when represented
as a set of time-ordered graphs, turns into six LF
time-ordered graphs, including two so-called ``stretched boxes'' with
two exchange particles in the intermediate state.  One of such
stretched boxes is shown in Fig.~\ref{feyn2}(b).  These stretched
boxes with two (and more) exchange particles in the intermediate state
are implicitly included in the BS equation, but they are not generated
by iteration of the LF ladder kernel; therefore, they are
omitted in the corresponding LFD bound state equation.  In principle,
this will cause a difference between the BS and LFD results.  However,
direct examination of the stretched boxes \cite{sbk} shows that they
are small.  Being explicitly included in the LF kernel, they
result in a small correction to the binding energy \cite{bs2}.  This
explains, why the BS and LFD two-body binding energies are very close
to each other, despite the fact that the kernels are not identical.

\begin{figure*}[htbp]
\begin{center}
\begin{minipage}{15.cm}
\mbox{\epsfxsize=6.5cm\epsffile{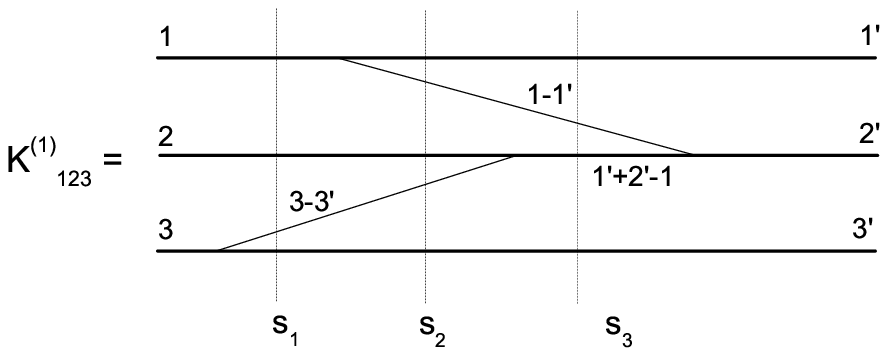} \hspace{0.5cm}
\epsfxsize=6.5cm\epsffile{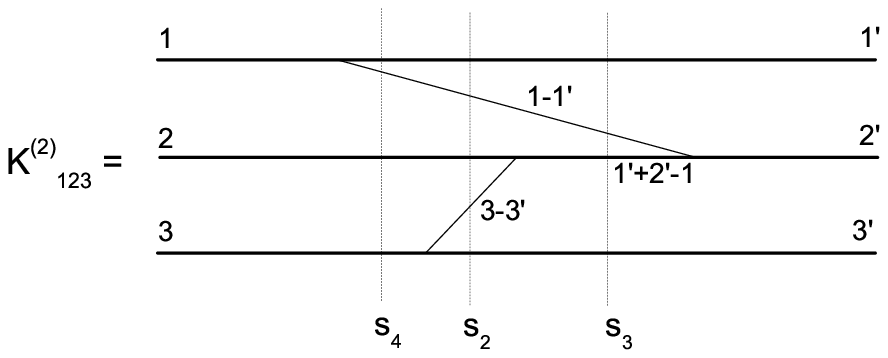}}
\end{minipage}
\end{center}
\begin{center}
\begin{minipage}{15.cm}
\mbox{\epsfxsize=6.5cm\epsffile{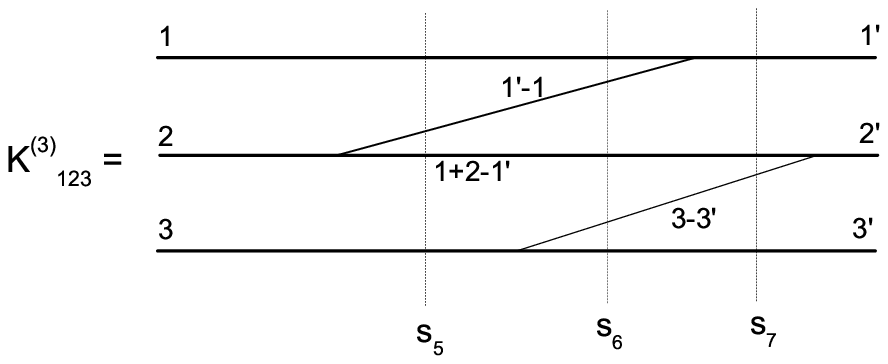} \hspace{0.5cm}
\epsfxsize=6.5cm\epsffile{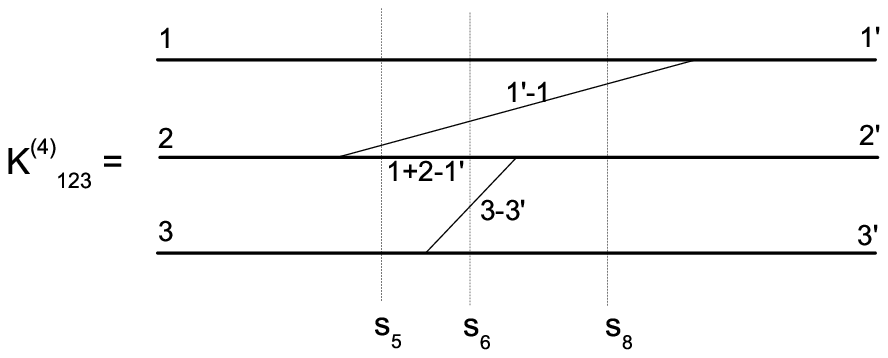}}
\end{minipage}
\end{center}
\begin{center}
\begin{minipage}{15.cm}
\mbox{\epsfxsize=6.5cm\epsffile{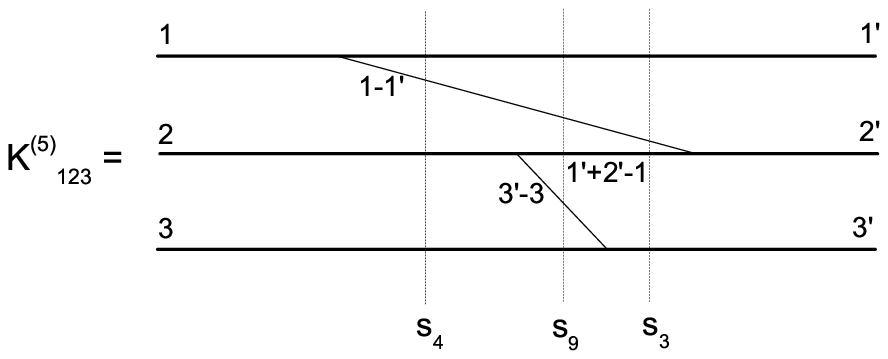} \hspace{0.5cm}
\epsfxsize=6.5cm\epsffile{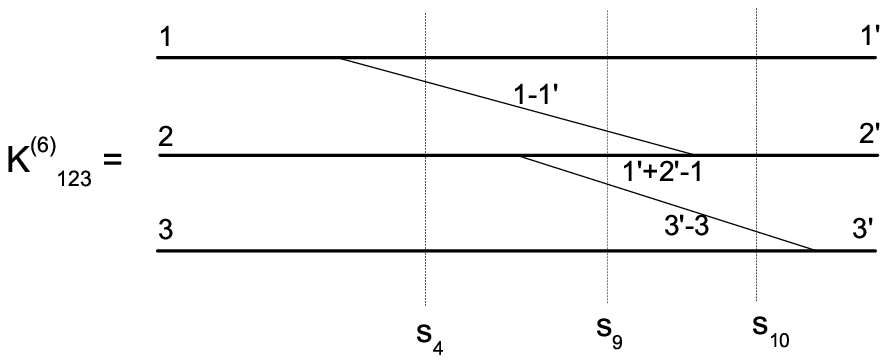}}
\end{minipage}
\end{center}
\begin{center}
\begin{minipage}{15.cm}
\mbox{\epsfxsize=6.5cm\epsffile{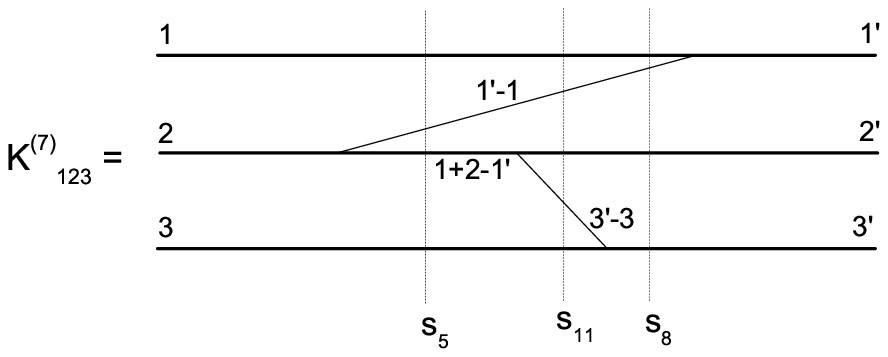} \hspace{0.5cm}
\epsfxsize=6.5cm\epsffile{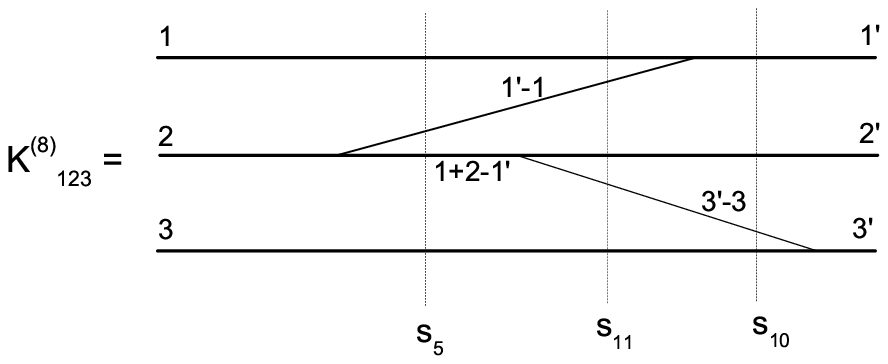}}
\begin{center}
\mbox{\epsfxsize=6.5cm\epsffile{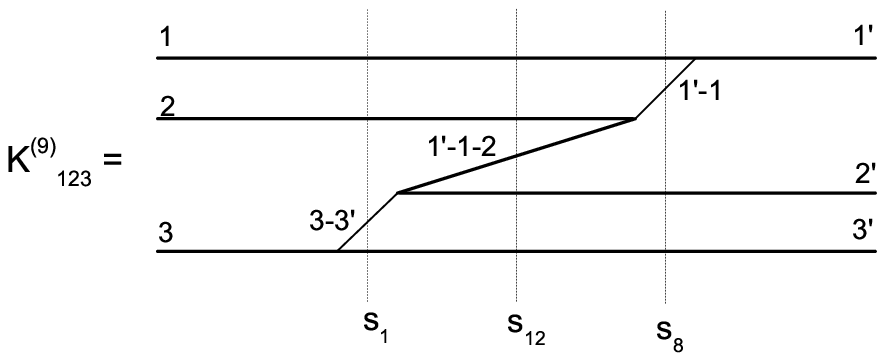}}
\end{center}
\end{minipage}
\end{center}
\caption{Contribution in three-body forces from the two
exchange-bosons in flight ($K^{(1-8)}_{123}$) and from pair-creation
($K^{(9)}_{123}$). \label{3bf}}
\end{figure*}

In the three-body problem, the situation is quite different: even with
a simple one-boson exchange kernel, new interesting and important
features emerge, highlighting differences between the BS approach and
LFD.  To be specific, in a three-body problem the difference between
LFD and time-ordered iterated BS kernels cannot be reduced to
stretched boxes: new LFD diagrams appear, which are shown in
Fig.~\ref{3bf}.
Like stretched boxes, these diagrams cannot be constructed by
iterations of the two-body LF kernel. In contrast to the stretched
boxes, they do not contain any loops. These graphs generate
three-body forces which are not taken into account in the
three-body LFD equation with ladder kernel. On the other hand,
they are implicitly included in the three-body BS equation.

Three-body nucleon forces have been discussed for several decades.
In recent years, there has been an increased interest in 3-nucleon
forces due to high precision few-body calculations. The need for
three-body forces results from nuclear phenomenology (see for
review Ref.~\cite{friar}).  In particular, three-body forces
remove the underbinding of light nuclei found with local two-body
forces.  With nonlocal two-body potentials there seems to be less
of a need for three-body forces from a phenomenological point of
view \cite{JISP16}. The diagrams depicted in Fig.~\ref{3bf}
represent a part of three-body forces for nucleons. Another part
arises due to excitation of intermediate isobars (most notably
$\Delta$-excitations).  The first-order relativistic corrections
due to the first eight graphs of Fig.~\ref{3bf}, was first found
in Ref.~\cite{brueckner}.  Its contribution to the triton binding
energy was calculated in Ref.~\cite{yang}.  In Ref.~\cite{glockle}
it was shown that this relativistic correction cancels with the
corresponding relativistic correction to the second iteration of
the one-boson exchange. Due to this cancellation, the sum of these
two corrections does not contribute in the Schr\"odinger equation.
However, in a truly relativistic framework, the full graphs
Fig.~\ref{3bf} (not a first-order relativistic correction) should
be taken into account.

Both the BS approach and LFD provide us with a relativistic
framework to study relativistic three-body bound states.  With
modern computer resources, we can now solve these bound state
equations in ladder truncation without further approximations (at
least for spinless systems).  These calculations serve as a
theoretical laboratory where we can study the importance of
three-body forces of relativistic origin, like the ones generated
by the diagrams in Fig.~\ref{3bf}.  By contrasting results from
the BS approach and LFD we can elucidate their contribution (if
any) to the binding energy. Note that in the scalar model under
consideration here, we have no analog of the three-body forces
originating from isobar excitations, so we cannot asses the
relative importance of three-body forces due to isobar
excitations; nor do we have any means to asses the role of
intrinsic three-body forces such as $NNN \to NNN$ point
interactions, or, in QCD, three-body forces due to the triple
gluon vertex.

The aim of this paper is two-fold: ({\it i}) First of all, we will
solve, for the first time, the three-body bound state BS and LFD
equations with a one-boson exchange kernel and compare the
corresponding binding energies. We shall see that, in contrast to the
two-body case, the binding energies found in these two approaches do
not coincide with each.  The LFD binding energy is significantly
smaller than the BS binding energy (i.e. the LF ladder kernel
underbinds compared to the BS ladder kernel). ({\it ii}) Then, in
order to identify explicitly the origin of this difference, we
calculate perturbatively the contribution to the LFD binding energy
from the three-body forces depicted in Fig.~\ref{3bf}.  It turns out
that this correction eliminates most of the difference.  In this way,
we see a clear and undoubted manifestation of the three-body forces
and establish that they are responsible for the underbinding of the
LF ladder kernel.

The plan of the paper is the following: In Sec.~\ref{BS} we
present the three-body BS equation, and we give a brief outline of
our numerical procedure.  The corresponding equation in LFD is
derived in Sec.~\ref{3beq}; the three-body kernel determined by
the graphs Fig.~\ref{3bf} and its perturbative contribution to the
binding energy are also given in this section.  Next, we present
our numerical calculations in Sec.~\ref{numeric}, where we
contrast the results in LFD with those obtained from the BS
equation.  Finally, Sec.~\ref{concl} contains concluding remarks.
Additional details regarding the relativistic LF Jacobi variables,
used in our calculations, can be found in Appendix~A. 

\section{Three-body Bethe--Salpeter equation
\label{BS}}
\begin{figure*}[tb]
\begin{center}
\includegraphics[width=12cm]{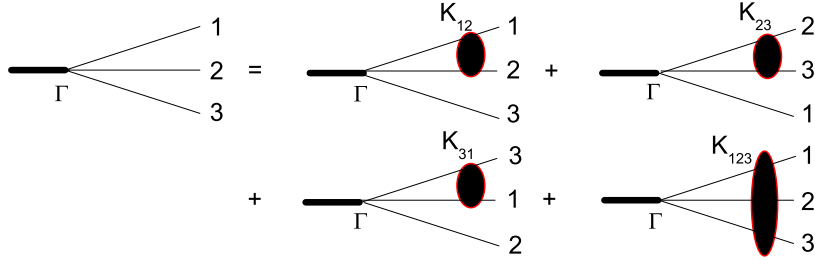}
\end{center}
\caption{Three-body bound state equation with two-body and three-body
forces.\label{fig1}}
\end{figure*}

A three-body bound state with total four-momentum $P$ can be
described by solution of the three-body bound state
equation\footnote{We use Euclidean metric in this section, so all
4-dimensional integrations are entirely Euclidean, and the
on-shell condition for the bound state implies $P^2 = -M^2$.}
\begin{eqnarray}
\Gamma(p_1,p_2,p_3; P) &=&
   \int\frac{d^4k_1}{(2\pi)^4}
   \int\frac{d^4k_2}{(2\pi)^4}
   \int\frac{d^4k_3}{(2\pi)^4} 
 \; \delta^4(P - \sum k_i)
\\ &\times& \!\!
   K(p_1,p_2,p_3; k_1,k_2,k_3; P) \;
  \Delta(k_1) \; \Delta(k_2) \;\Delta(k_3) \; \Gamma(k_1,k_2,k_3; P)
  \nonumber
\end{eqnarray}
at $P^2 = -M^2$, where $M$ is the three-body bound state mass. Here,
$K$ is the three-body scattering kernel, which can be decomposed into
three 2-body kernels $K_{ij}$ and an intrinsic 3-body kernel
$K_{123}$, see Fig.~\ref{fig1}; $\Delta$ are the (dressed) propagators
for the constituent particles.
Momentum conservation dictates $p_1 + p_2 + p_3 = k_1 + k_2 + k_3 =
P$, which is enforced (on the right-hand-side) by the
$\delta$-function.  In the absence of intrinsic three-body kernels
($K_{123}$ in Fig.~\ref{fig1}), the three-body bound state equation
reduces to
\begin{eqnarray}
 \Gamma(p_1,p_2,p_3; P)  & = &
   \int\frac{d^4k}{(2\pi)^4}
   K_{12}(p_1,p_2; k_1,\tilde{k}_2; P) \;
   \Delta(k_1) \; \Delta(\tilde{k}_2) \; \Gamma(k_1,\tilde{k}_2,p_3; P)
\nonumber \\ &+&
   \int\frac{d^4k}{(2\pi)^4}
   K_{23}(p_2,p_3; k_2,\tilde{k}_3; P) \;
   \Delta(k_2) \; \Delta(\tilde{k}_3) \; \Gamma(p_1,k_2,\tilde{k}_3; P)
\nonumber \\ &+&
   \int\frac{d^4k}{(2\pi)^4}
   K_{31}(p_3,p_1; k_3,\tilde{k}_1; P) \;
\Delta(k_3) \; \Delta(\tilde{k}_1) \; \Gamma(\tilde{k}_1,p_2,k_3;
P),
\nonumber\\
&&
\end{eqnarray}
where $k_i = p_i + k$ and $\tilde{k}_j = p_j - k$ and with $K_{ij}$
the same two-body kernels as in the two-body BS equation.

\subsection{Scalar three-body BS equation}
For simplicity, we consider a scalar field theory, with
interaction $-g\phi^2 \varphi$, and a three-body bound state
consisting of three particles $\phi$, interacting via the exchange
of a $\varphi$ boson. We take equal mass $m$ for the constituents,
and assume that they are all distinguishable; the mass of the
exchange boson is $\mu$.  In the limit $\mu = 0$ this becomes the
well-known Wick--Cutkosky model \cite{WickCutkosky}.

We use the ladder truncation for the three-body bound state equation.
Furthermore, we do not take into account any self-energies (i.e. we
use bare propagators).  With these truncations, the three-body BS
equation reduces to (see also Fig.~\ref{fig1})
\begin{eqnarray}
\Gamma(p_1, p_2, p_3;P) \;=\; g^2 \sum
 \int\frac{d^4k}{(2\pi)^4}  \; D(k) \;
  \Delta(k_1) \; \Delta(\tilde{k}_2) \;
      \Gamma(k_1,\tilde{k}_2,p_3;P),
\end{eqnarray}
where $k_i = p_i + k$, $\tilde{k}_j = p_j - k$, and the sum runs
(cyclic) over $i,j = 1, 2, 3$, $i \neq j$.  The propagators are:
scalar constituent propagator
\begin{eqnarray*}
    \Delta(p) &=& \frac{1}{p^2 + m^2}\, ,
\end{eqnarray*}
and scalar exchange propagator
\begin{eqnarray*}
    D(k) &=& \frac{1}{k^2 + \mu^2}\, .
\end{eqnarray*}

The momenta satisfy momentum conservation, so the three momenta $p_i$
are not independent.  The bound state amplitude $\Gamma$ is actually
function of only two (relative) independent momenta, and one can write
it as a function of only two independent 4-vectors
\begin{eqnarray*}
  \Gamma(p_1, p_2, p_3;P) &=& \Gamma(p, q;P).
\end{eqnarray*}
One can make different choices for these two relative momenta $p$ and
$q$; the physics should be independent of these choices.  The most
natural choice are the Jacobi variables in momentum space
\begin{eqnarray*}
   p_1 &=& \case{1}{3} P + q + p,
\\
   p_2 &=& \case{1}{3} P + q - p,
\\
   p_3 &=& \case{1}{3} P - 2\,q.
\end{eqnarray*}
However, this choice is not unique, and we have used other choices as
well; within the estimated numerical errors, our results are
independent of this choice (as it should be).

In this notation, the ladder BS equation for $\Gamma(p,q; P)$
becomes
\begin{eqnarray}\label{Eq:bse}
\Gamma(p,q;P) &=& g^2 \int\frac{d^4k}{(2\pi)^4} \; D(k)  
\\ &\times&  \Big[\Delta\left(\case{1}{3}P + q + p +
k\right) \;
  \Delta\left(\case{1}{3}P + q - p - k\right) \;
   \Gamma(p+k,q;P)
\nonumber \\
 &+&  \phantom{\Big[}\Delta\left(\case{1}{3}P + q - p +
k\right) \;
  \Delta\left(\case{1}{3}P - 2 q - k\right) \;
      \Gamma(p-\case{1}{2}k,q+\case{1}{2}k;P)
\nonumber \\
 &+&  \phantom{\Big[}\Delta\left(\case{1}{3}P - 2 q + k\right)
\;
  \Delta\left(\case{1}{3}P + q + p - k\right) \;
      \Gamma(p-\case{1}{2}k,q-\case{1}{2}k;P)
\Big]. \nonumber
\end{eqnarray}
The momentum of the exchanged particle, namely the integration
variable $k$, is the same in all three terms.

\subsection{Numerical implementation}
As mentioned, the bound state amplitude $\Gamma$ is a function of
two independent four-vectors $p$ and $q$, but the actual
independent variables are: two radial variables, $p^2$ and $q^2$,
and three angles, $\theta_p$, $\theta_q$, and $\phi_{pq}$.  In the
rest-frame of the bound state, $P_\mu = [i \, M, 0, 0, 0]$, we use
the notation
$$
\begin{array}{l}
 p_\mu \; = \; p\, [\cos(\theta_p), \sin(\theta_p), 0, 0]
\\
 q_\mu \; = \; q\, [\cos(\theta_q), \sin(\theta_q)\cos(\phi_{pq}),
          \sin(\theta_q)\sin(\phi_{pq}), 0]
\end{array}
$$
and we have the 4-dimensional integration measure
$$
 \int d^4k =  \int_0^\infty k^3 \; dk
    \int_0^\pi \sin^2(\theta) \; d\theta
    \int_0^\pi \sin(\phi) \; d\phi
    \int_0^{2\pi} d\gamma\, .
$$
All three angular integrations have to be done numerically, in
addition to the radial integration.  Furthermore, we need to do
some interpolation on the function $\Gamma(p,q;P)$ we are solving
for, because we have expressions like $\Gamma(p - \case{1}{2}
k,q\pm \case{1}{2} k;P)$ under the integral. Thus we have a
four-dimensional integral equation for the bound state amplitude
which is a function of five independent variables, given a fixed
(external) $P^2 = -M^2$, with a nine-dimensional kernel.  For a
given choice of $P^2 = -M^2$ (or of the binding energy $E_b = 3 \,
m - M$), we can solve this (four-dimensional) eigenvalue equation
for the corresponding coupling constant $g$, or rather, for the
dimensionless constant $\alpha=g^2 / (16 \pi m^2)$; the
(five-dimensional) eigenvector is the BS amplitude,
$\Gamma(p^2,q^2,\theta_p,\theta_q,\phi_{pq};P^2)$.

We solve this numerically by straightforward discretization of the
internal and external variables; we do not make use of any
expansion of $\Gamma$ in partial waves, nor any other set of basis
functions. No matter how we re-arrange the internal and external
variables, we have to do some interpolation on the function
$\Gamma$ we want to solve for; that means that there is nothing to
be gained by taking the internal and external grids to be the
same.  After a suitable transformation $q^2 \to x$, with $0 < q^2
< \infty$, and $0 < x < 1$, we split the radial interval into $N$
sub-intervals, and use 2-point Gaussian integration inside each
sub-interval; for the external radial grid we use the $N$
break-points (with the boundary condition that for $x=1$, i.e.
$q^2 \to \infty$ the BS amplitude vanishes).  For the angular
integrations we use suitable Gaussian integration measures:
Gauss--Chebyshev for $\cos\theta$, Gauss--Legendre for $\cos\phi$
and $\gamma$.  For the external angular variables
$0<\theta_p<\pi$, $0<\theta_q<\pi$, and $0<\phi_{pq}<\pi$ and we
use equidistant grids in $\cos\theta_{p,q}$ and $\cos\phi_{pq}$.

The ground state solution (that is all we are interested in right
now) is symmetric under exchange of any of the three constituents.
That means that $\Gamma(p,q; P) = \Gamma(-p,q; P)$, i.e. the
solution we are looking for is symmetric under the combined
transformation
$$
 \cos(\theta_p) \; \rightarrow \; -\cos(\theta_p),
\quad
 \cos(\phi_{pq}) \; \rightarrow \; -\cos(\phi_{pq}).
$$
Using this symmetry property saves a factor of two in the
computational effort.

\begin{figure}[tb]
\includegraphics[width=0.96\columnwidth]{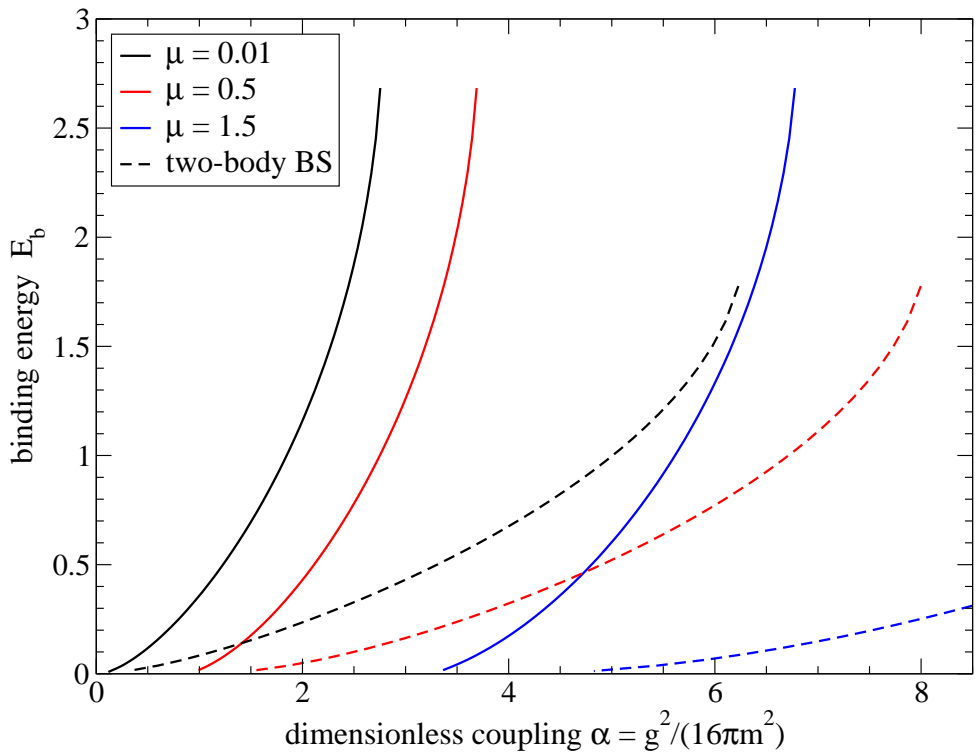}
\caption{(Color online) Binding energies obtained by solution of the
BS equation in ladder truncation for three different values of the
exchange-boson mass: $\mu=0.01$, $\mu=0.5$, and $\mu=1.5$ (units set
by $m
= 1$).  Solid curves are the
three-body BS result; dashed curves are the two-body BS
results.\label{figBSE2E3}}
\end{figure}
Since we are only interested in the largest eigenvalue, and its
corresponding eigenvector (i.e. in the ground state), we can solve the
eigenvalue equation by iteration (also called power method).
Typically, we need of the order of 25 to 40 iterations in order to
obtain a stable solution.  Our results are shown in
Fig.~\ref{figBSE2E3} for three different values of the mass of the
exchange boson.  The mass $m$ of the constituents is set to unity, so
the maximal binding energy for a three-body system is three.  For
comparison, we also show the binding energy of the two-body system
(also in ladder truncation), which clearly shows that the three-body
system is stronger bound than the corresponding two-body system;
i.e. the solutions we find are indeed three-body bound states.

\section{Three-body bound state equation in LFD}\label{3beq}
The most simple way to derive the LF equation for spinless
particles and calculate the kernel is to use the Weinberg rules
\cite{sw}, which are equivalent to the graph techniques in LFD
\cite{cdkm,bpp}.  To calculate the amplitude $-{\cal M}$, one
should put in correspondence:
\begin{itemize}
\item to every vertex -- the factor $g$,
\item to every intermediate state -- the factor
 $$ \frac{2}{s_0-s_{int}+i0},\quad \mbox{where}\quad
    s_{int}=\sum_i \frac{\vec{k}^2_{i\perp}+m_i^2}{x_i},
 $$
and $s_0$ is the initial (=final) state energy.
\end{itemize}
In our case (three-body bound state): $s_0=M_3^2$, where $M_3$ is mass
of the bound state three-body system.  To every internal line one
should put in correspondence the factor $\frac{\theta(x_i)}{2x_i}$.
One should take into account the conservation laws for
$\vec{k}_{i\perp}$ and $x_i$ in any vertex and integrate over all
independent variables with the measure
$\frac{d^2k_{i\perp}dx_i}{(2\pi)^3}$.

Applying these rules to the graph shown in Fig.~\ref{fig1}, we find
\begin{eqnarray*}
\Gamma &=&  
\int \frac{2\; \Gamma\, K_{12}}
{s'_{123}-M_3^2}\,\frac{\theta(x'_1)}{2x'_1}\frac{\theta(x'_2)}{2x'_2}\,
\frac{d^2{k'}_{1\perp}dx'_1}{(2\pi)^3}
 + (231) + (312)
\\ & +&\sum
 \int \!\!\frac{2\,\Gamma\,K_{123}}{s''_{123}-M_3^2}
 \frac{\theta(x'_1)}{2x'_1} \frac{\theta(x'_2)}{2x'_2}
 \frac{\theta(x'_3)}{2x'_3}
 \frac{d^2{k'}_{1\perp}dx'_1}{(2\pi)^3}
 \frac{d^2{k'}_{3\perp}dx'_3}{(2\pi)^3}\, .
\end{eqnarray*}
Here $x'_1+x'_2+x_3=1$ in the first term, and $x'_1+x'_2+x'_3=1$
in the last term, $s'_{123}$ is the three-body energy in the
intermediate state, with two interacting particles 12, whereas the
particle 3 is free (first graph in r.h.s. of Fig.~\ref{fig1});
$s''_{123}$ is the three-body energy in the intermediate state
with all three interacting particles (last graph in r.h.s. of
Fig.~\ref{fig1}), and $\Gamma$ is the LF vertex function.  The
corresponding LF wave function reads
$$
\psi(1,2,3)=\frac{\Gamma(1,2,3)}{s_{123}-M_3^2},
$$
where
\begin{equation}\label{s123}
s_{123}=\frac{k_{1\perp}^2+m^2}{x_1}+\frac{k_{2\perp}^2+m^2}{x_2}+
\frac{k_{3\perp}^2+m^2}{x_3}
\end{equation}
with $\vec{k}_{1\perp}+\vec{k}_{2\perp}+\vec{k}_{3\perp}=0$ and
sum of all $x$'s is 1, as always. $\Gamma(1,2,3)$ means:
$\Gamma(1,2,3)=
\Gamma(\vec{k}_{1\perp},x_1;\vec{k}_{2\perp},x_2;\vec{k}_{3\perp},x_3)$
and similarly for $\psi(1,2,3)$.

Now we will use the LF Jacobi momenta
$\vec{k}_{\perp}\equiv\vec{k}_{12\perp},x\equiv x_{12}$ constructed in
Appendix A.
I.e., we make the substitution:
\begin{equation}\label{kx}
\begin{array}{ll}
\vec{k}_{1\perp}=\vec{k}_{\perp}-x\vec{k}_{3\perp},& x_1=x(1-x_3),
\\
\vec{k}_{2\perp}=-\vec{k}_{\perp}-(1-x)\vec{k}_{3\perp},&
x_2=(1-x)(1-x_3),
\\
\vec{k'}_{1\perp}=\vec{k'}_{\perp}-x'\vec{k'}_{3\perp},&
x'_1=x'(1-x'_3),
\\
\vec{k'}_{2\perp}=-\vec{k'}_{\perp}-(1-x')\vec{k'}_{3\perp},&
x'_2=(1-x')(1-x'_3).
\end{array}
\end{equation}
For $\psi$ (and $\Gamma$) we get $\psi(1,2,3)=
\psi(\vec{k}_{\perp},x;\vec{k}_{3\perp},x_3)$.  Now the variables
$\vec{k}_{\perp},x;\vec{k}_{3\perp},x_3$ are independent and they
are not constrained by any relations.  For both $x$'s we have:
$0\leq x \leq 1$, $0\leq x_3 \leq 1$.  Then $s_{123}$,
Eq.~(\ref{s123}), becomes
\begin{equation}\label{s123p}
s_{123}= \frac{k_{3\perp}^2+s_{12}}{1-x_3}
+\frac{k_{3\perp}^2+m^2}{x_3}\ ,
\end{equation}
where
$$
s_{12}=\frac{k_{\perp}^2+m^2}{x(1-x)}
$$ is the effective mass squared of two-body subsystem.  The effective
three-body mass squared $s_{123}$ has now the form of an effective
mass of a two-body system, consisting of masses $\sqrt{s_{12}}$ and
$m$.

In terms of Jacobi momenta, the equation for $\psi$ becomes
\begin{eqnarray}
(s_{123}-M_3^2)\;\psi &=&
\int \psi\, K_{123}\;
\frac{d^2{k'}_{\perp}dx'}{(2\pi)^3 2x'(1-x')}\;
\frac{d^2{k'}_{3\perp}dx'_3}{(2\pi)^32x'_3(1-x'_3)} \nonumber \\
&-&\frac{m^2}{2\pi^3}\int \psi\, V_{12}\;
\frac{d^2{k'}_{\perp}dx'}{(2\pi)^32 x'(1-x')(1-x_3)} \nonumber \\
&-& (231) - (312)\ . \label{eq3b}
\end{eqnarray}
Here, instead of $K_{12}$, given by the Weinberg rules, we have
introduced the two-body kernel $V_{12}=-\frac{1}{4m^2}K_{12}=
V(\vec{k'}_{\perp},x';\vec{k}_{\perp},x;M^2)$, which in the
non-relativistic limit turns into the potential in the
Shr\"odinger equation. The kernel
$V(\vec{k'}_{\perp},x';\vec{k}_{\perp},x;M^2)$ has exactly the
two-body form, i.e. the form which enters the two-body LF
equation. For one-boson exchange it is given in the next section.

\subsection{Scalar LF bound state equation}
In ladder truncation, there is no intrinsic three-body kernel, so
we omit the first term in Eq.~(\ref{eq3b}) for now.  However, we
will find the three-body kernel $K_{123}$ due to the diagrams of
Fig.~\ref{3bf} in Sec.~\ref{3bk} below, and calculate its
contribution to the binding energy perturbatively in the next
section.  After omitting $K_{123}$, the LF bound state equation,
Eq.~(\ref{eq3b}), contains three interaction kernels
$V_{12},V_{23},V_{31}$.

For the one-boson exchange model with mass $\mu$ the two-body
potential reads (see e.g. \cite{cdkm}):
\begin{eqnarray}\label{kernrx}
&&V(\vec{k'}_{\perp},x';\vec{k}_{\perp},x;M^2)=
\\
&&-\frac{4\pi\alpha}{\mu^2+\frac{x'}{x}\left(1-\frac{x}{x'}\right)^2m^2
+\frac{x'}{x}\left(\vec{k}_{\perp}-\frac{x}{x'}\vec{k'}_{\perp}\right)^2
+(x'-x)\left(\frac{m^2+{\vec{k'}_{\perp}}^2}{x'(1-x')}-M^2\right)}\
. \nonumber
\end{eqnarray}
This expression is valid for $x \leq x'$. If $x \geq x'$, the
kernel is obtained from (\ref{kernrx}) by the permutation
$x\leftrightarrow x'$, $\vec{k}_{\perp}\leftrightarrow
\vec{k'}_{\perp}$. Here $\alpha$ is the same dimensionless
coupling constant as introduced in the previous section,
$\alpha={g^2}/{(16\pi m^2)}$, with $g$ the coupling constant in
the Hamiltonian $H=-g\phi^2\varphi$. In the non-relativistic limit
the kernel, Eq.~(\ref{kernrx}), turns into the well-known Yukawa
potential
\begin{equation}\label{yukawa}
V(\vec{k}-\vec{k'})=\frac{-4\pi\alpha}{\mu^2+(\vec{k}-\vec{k'})^2}
\end{equation}
which in the coordinate space is simply
$$
 V(r)=-\frac{\alpha e^{-\mu r}}{r} \;.
$$

Now we introduce the Faddeev components
\begin{eqnarray}\label{fad}
\psi(1,2,3)&=&\psi_{12}(1,2,3)+\psi_{23}(1,2,3)+\psi_{31}(1,2,3)
\nonumber\\
&=&\psi_{12}(1,2,3)+\psi_{12}(2,3,1)+\psi_{12}(3,1,2) \,,
\end{eqnarray}
such that the component, for instance, $\psi_{12}(1,2,3)$ satisfies an
equation containing $V_{12}$ only.  This procedure is standard.  In
this way we obtain the equation for the component $\psi_{12}$
\begin{eqnarray}\label{eq21}
&&(s_{123}-M_3^2)\,\psi_{12}(\vec{k}_{\perp},x;\vec{k}_{3\perp},x_3) \;=\;
\\ && \phantom{(s_{123}-M_3^2)\,}
-\frac{\displaystyle{m^2}}{\displaystyle{2\pi^3}}\,
\displaystyle{\int} \frac{\displaystyle{d^2k'_{\perp}dx'}}
{\displaystyle{2x'(1-x')}}
\;\;\frac{\displaystyle{1}}{\displaystyle{(1-x_3)}}
\;\;V(\vec{k}_{\perp},x;\vec{k'}_{\perp},x';M^2_{12}) \cr
\nonumber\\ & \times& \Big[
\psi_{12}\left(\vec{k'}_{\perp},x';\vec{k}_{3\perp},x_3\right)
+\psi_{12}\left(\vec{k'}_{23\perp},x'_{23};\vec{k'}_{1\perp},x'_1\right)
+ \psi_{12}\left(\vec{k'}_{31\perp},x'_{31};
\vec{k'}_{2\perp},x'_2\right)\Big]\ ,
 \nonumber
\end{eqnarray}
where $s_{123}$ is defined by Eq.~(\ref{s123p}) and
\begin{equation}\label{eq17}
{M}^2_{12}=(1-x_3)M_3^2- \displaystyle{
\frac{k_{3\perp}^2+(1-x_3)m^2}{x_3}}\ .
\end{equation}
The three-body mass $M_3^2$ (which we are solving for) enters this
equation through the factor $(s_{123}-M_3^2)$ in l.h.s. and through
the value of $M_{12}^2$, Eq.~(\ref{eq17}), in an argument of $V$.  (An
equation similar to Eq.~(\ref{eq21}), but for zero range interaction,
was firstly derived and solved numerically in
Refs.~\cite{tobias,ck_03}.)

The second and third terms in Eq.~(\ref{eq21}), namely
$\psi_{12}\left(\vec{k'}_{23\perp},x'_{23};\vec{k'}_{1\perp},x'_1\right)$
and
$\psi_{12}\left(\vec{k'}_{31\perp},x'_{31};\vec{k'}_{2\perp},x'_2\right)$
depend on the LF Jacobi variables (23,1) and (31,2) respectively,
whereas the equation is written in terms of the variables (12,3).  The
variables (23,1) and (31,2) should be expressed though (12,3).  These
expressions are derived in Appendix \ref{permut}.  In particular, the
variables $\vec{k'}_{23\perp},x'_{23},\vec{k'}_{1\perp},x'_1$ and
$\vec{k'}_{31\perp},x'_{31},\vec{k'}_{2\perp},x'_2$ are obtained from
Eqs.~(\ref{v23}) and (\ref{v31}) of Appendix \ref{permut} by the
replacement: $\vec{k}_{\perp}\to\vec{k'}_{\perp}$, $x\to x'$.  To be
specific:
\begin{equation}\label{v23p}
\begin{array}{ll}
\vec{k}'_{23\perp}=\displaystyle{-\frac{x_3\vec{k}'_{\perp}
+(1-x')\vec{k}_{3\perp}} {1-x'+x'x_3}},&
x'_{23}=\displaystyle{\frac{(1-x')(1-x_3)}{1-x'+x'x_3}},
\\
&
\\
\vec{k'}_{1\perp}=\vec{k'}_{\perp}-x'\vec{k}_{3\perp},&
x'_1=x'(1-x_3),
\end{array}
\end{equation}
and
\begin{equation}\label{v31p}
\begin{array}{ll}
\vec{k'}_{31\perp}=\displaystyle{-\frac{x_3\vec{k'}_{\perp}
-x'\vec{k}_{3\perp}}{x'+x_3-x'x_3}},&
x'_{31}=\displaystyle{\frac{x_{3}}{x'+x_3-x'x_3}},
\\
&\\
\vec{k'}_{2\perp}=-\vec{k'}_{\perp}-(1-x')\vec{k}_{3\perp},&
x'_2=(1-x')(1-x_3).
\end{array}
\end{equation}
With this notation (and omitting prime), the full wave function
obtains the form:
\begin{eqnarray}\label{psitot}
\psi(\vec{k}_{\perp},x;\;\vec{k}_{3\perp},x_3) &=&
\psi_{12}(\vec{k}_{\perp},x;\;\vec{k}_{3\perp},x_3)  \\
&+& \psi_{12}(\vec{k}_{23\perp},x_{23};\;\vec{k}_{1\perp},x_1) +
\psi_{12}(\vec{k}_{31\perp},x_{31};\;\vec{k}_{2\perp},x_2)\ .
\nonumber
\end{eqnarray}
It is normalized as
\begin{equation}\label{norm}
\int \big|\psi(\vec{k}_{\perp},x;\vec{k}_{3\perp},x_3)\big|^2\;
\frac{d^2k_{\perp}dx}{(2\pi)^32x(1-x)}\;
\frac{d^2k_{3\perp}dx_3}{(2\pi)^32x_3(1-x_3)} \;=\; 1\ .
\end{equation}

\subsection{Contribution from three-body kernel
\label{3bk}}
Naively, one might not expect any three-body kernels in the ladder
truncation.  However, in LFD, there are three-body kernels, as
already mentioned in the introduction, due to the (LF)
time-ordering of the diagrams.  To find the correction to the
binding energy from the kernel $K_{123}$, we represent equation
for the full wave function symbolically in the form
\begin{eqnarray*}
(s_{123}-M_3^2)\psi &=&
(\hat{K}_{12}+\hat{K}_{23}+\hat{K}_{31})\psi+ \hat{K}_{123}\psi \,.
\end{eqnarray*}
Substituting $M_3^2 \to M_3^2+\Delta M_3^2$, we find
\begin{eqnarray*}
\Delta M_3^2 &=& -(\psi \hat{K}_{123}\psi),
\end{eqnarray*}
or, explicitly
\begin{eqnarray}\label{E30}
&&\Delta M^2_3  = \; -
\int\frac{d^2k_{3\perp}}{(2\pi)^3}\;\frac{d^2k_{\perp}}{(2\pi)^3}\;
\frac{d^2k'_{3\perp}}{(2\pi)^3}\;\frac{d^2k'_{\perp}}{(2\pi)^3} 
\\ &&\times \int_0^1 \frac{dx_3}{2x_3(1-x_3)}\;
\frac{dx}{2x(1-x)} \; \frac{dx'_3}{2x'_3(1-x'_3)} \;
\frac{dx'}{2x'(1-x')} \nonumber \\ &&\times
 \psi(\vec{k}_{\perp},x;\vec{k}_{3\perp},x_3)\,
 K_{123}(\vec{k}_{\perp},x;\vec{k}_{3\perp},x_3;
         \vec{k'}_{\perp},x';\vec{k'}_{3\perp},x'_3)\,
\psi(\vec{k'}_{\perp},x';\vec{k'}_{3\perp},x'_3) \,. \nonumber
\end{eqnarray}
The integrand depends on four 2-dimensional vectors $\vec{k}_{\perp}$,
$\vec{k}_{3\perp}$, $\vec{k'}_{\perp}$, $\vec{k'}_{3\perp}$ and,
hence, on three relative angles between them.  Therefore
Eq.~(\ref{E30}) is an 11-dimensional integral (the last, fourth angle
integration, is reduced to a factor $2\pi$).

The kernel $K_{123}$ is determined by 54 graphs: 48 of them contain
two exchange-bosons in flight, whereas another 6 correspond to the
intermediate production of a pair of constituent
particle-antiparticles.  It is sufficient to calculate only the nine
graphs shown in Fig.~\ref{3bf}.  The other contributions are obtained
by permutations and can be taken into account by multiplying the
results of these nine graphs by a factor 3!=6.

Here we show explicitly how to calculate the contribution of the
first graph Fig.~\ref{3bf} only.  All other contributions are
calculated in a similar fashion.  Following the Weinberg rules
\cite{sw}, we get
\begin{equation}\label{K1}
K^{(1)}_{123} =\frac{2^8\pi^2 m^4\alpha^2}
{(s_1-M_3^2)\,(s_2-M_3^2)\, (s_3-M_3^2)}
\frac{\theta(x_1-x'_1)}{(x_1-x'_1)}\,
\frac{\theta(x'_1+x'_2-x_1)}{(x'_1+x'_2-x_1)}\,
\frac{\theta(x_3-x'_3)}{(x_3-x'_3)} \,,
\end{equation}
where $s_{i}$ are the energies in intermediate states shown in
Fig.~\ref{3bf}.  The line 1, for instance, is associated with the
momenta $\vec{k}_{1\perp},x_1$ and contributes in $s_1$ the term
$\frac{k_{1\perp}^2+m^2}{x_1}$, and analogously for other lines.
We sum over all the lines in given intermediate state. In this way
we find for $s_1$:
$$
s_1 = \frac{k_{1\perp}^2+m^2}{x_1}+\frac{k_{2\perp}^2+m^2}{x_2}
+\frac{(\vec{k}_{3\perp}-\vec{k'}_{3\perp})^2+\mu^2}{x_3-x'_3}+
\frac{{k'}^2_{3\perp}+m^2}{x'_3} \,.
$$
Other intermediate state energies can be found similarly.
Transforming this expression to the LF Jacobi variables by the
substitution (\ref{kx}), we find:
$$
s_1 =
\frac{(\vec{k}_{3\perp}-\vec{k'}_{3\perp})^2+\mu^2}{x_3-x'_3}
+\frac{\vec{k}\,^2_{\perp}+x(1-x)\vec{k}\,^2_{3\perp}+m^2}{x(1-x)(1-x_3)}
+\frac{\vec{k'}\,^2_{3\perp}+m^2}{x'_3} \,.
$$

The final expression for the kernel $K^{(1)}_{123}$ is obtained by
substitution of this expression for $s_1$ (and analogously, for
$s_2,s_3$) in Eq.~(\ref{K1}) and replacing also the variables
$x$'s in Eq.~(\ref{K1}) by the Jacobi ones according to
Eqs.~(\ref{kx}). The $\theta$-functions impose the following
constrains on the Jacobi's $x$'s:
\begin{eqnarray*}
 0\leq x_3 \leq 1,\qquad  0\leq x \leq 1, \qquad 0 \leq x'_3 \leq x_3,
\end{eqnarray*}
and
\begin{eqnarray*}
 0\leq x' \leq \frac{x(1-x_3)}{1-x'_3} \,.
\end{eqnarray*}

Other contributions are found similarly.  All of them have the
form (\ref{K1}) with the replacement of $s_1,s_2,s_3$ by the
corresponding twelve intermediate energies $s_1-s_{12}$, as shown
in the various graphs in Fig.~\ref{3bf}, and also $x$'s in the
$\theta$-functions and in denominator. Taking the sum of all the
contributions and substituting it in Eq.~(\ref{E30}), we find the
perturbative correction to the binding energy resulting from the
three-body forces:
\begin{eqnarray} \label{DM32}
\Delta M_3^2 &=& - 6
\int\frac{d^2k_{3\perp}}{(2\pi)^3}\;\frac{d^2k_{\perp}}{(2\pi)^3}\;
\frac{d^2k'_{3\perp}}{(2\pi)^3}\;\frac{d^2k'_{\perp}}{(2\pi)^3}
\int_0^1 \frac{dx_3}{2x_3(1-x_3)} \int_0^1 \frac{dx}{2x(1-x)}
\nonumber\\
&\times& \Bigg\{ \int_0^{x_3} \frac{dx'_3}{2x'_3(1-x'_3)}
\nonumber\\ & \times& \Bigg[\int_0^{\frac{x(1-x_3)}{1-x'_3}}
\frac{\psi (K^{(1)}_{123}+K^{(2)}_{123}) \psi'\;dx'}{2x'(1-x')} +
\int_{\frac{x(1-x_3)}{1-x'_3}}^{\frac{1-x_3}{1-x'_3}} \frac{\psi
(K^{(3)}_{123}+K^{(4)}_{123})\psi'\;dx'}{2x'(1-x')} \Bigg]
\nonumber\\
&+&\phantom{\Bigg\{} \int_{x_3}^{1-x(1-x_3)} \!\!\!\!\!\!
 \frac{dx'_3}{2x'_3(1-x'_3)}
\nonumber\\ &\times&\Bigg[ \int_0^{\frac{x(1-x_3)}{1-x'_3}}
\frac{\psi (K^{(5)}_{123}+K^{(6)}_{123}) \psi'\;dx'}{2x'(1-x')} +
\int_{\frac{x(1-x_3)}{1-x'_3}}^{1} \frac{\psi
(K^{(7)}_{123}+K^{(8)}_{123}) \psi'\;dx'}{2x'(1-x')}\Bigg]
\nonumber\\
&+& \int_0^{x_3} \frac{dx'_3}{2x'_3(1-x'_3)}
\int_{\frac{1-x_3}{1-x'_3}}^{1} \frac{\psi K^{(9)}_{123}
\psi'\;dx'}{2x'(1-x')} \Bigg\}\ .
\end{eqnarray}
The three-body wave function $\psi$ is expressed through the Faddeev
component $\psi_{12}$ by Eq.~(\ref{psitot}), whereas $\psi_{12}$ is
found from Eq.~(\ref{eq21}); $\psi$ and $\psi'$ depend on the
variables $\vec{k}_{\perp},x;\;\vec{k}_{3\perp},x_3$ and
$\vec{k'}_{\perp},x';\;\vec{k'}_{3\perp},x'_3$ respectively.

\section{Numerical results
\label{numeric}}
The LF wave functions in Eq.~(\ref{eq21}) depend on five
independent scalar variables\footnote{We remind that three-body BS
amplitude also depends on five scalar variables. This coincidence
takes place for two- and three-body systems only. For $n$-body
system with $n\geq 4$ LF wave function and BS amplitude depend on
different numbers of scalar variables.}.  The wavefunction
$\psi_{12}(\vec{k}_{\perp},x;\vec{k}_{3\perp},x_3)$ on the l.h.s.
of Eq.~(\ref{eq21}) depends on the external variables $k_{\perp}$,
$k_{3\perp}$, $x$, $x_3$, and
$z=\cos\beta=\vec{k}_{\perp}\cdot\vec{k}_{3\perp}/(k_{\perp}k_{3\perp})$.
The arguments of the wave functions $\psi$'s in the integrand on
the r.h.s. of Eq.~(\ref{eq21}) can be expressed through a
combination of these five external variables and the three
integration variables $k'_{\perp}$, $x'$, and $\cos\beta'=
\vec{k}_{\perp}\cdot\vec{k'}_{\perp}/(k_{\perp}k'_{\perp})$. Thus
the structure of the LF bound state equation, Eq.~(\ref{eq21}), is
very similar to that of the BS equation, Eq.~(\ref{Eq:bse}), and
we can solve it in a similar fashion.  That is, we use a
transformation to map the radial variable onto a finite interval,
discretize all five variables, and perform all integrations
numerically (three in the case of LFD, four in the case of the BS
equation).

\subsection{Difference between BS approach and LFD}
\begin{figure}[tb]
\includegraphics[width=0.96\columnwidth]{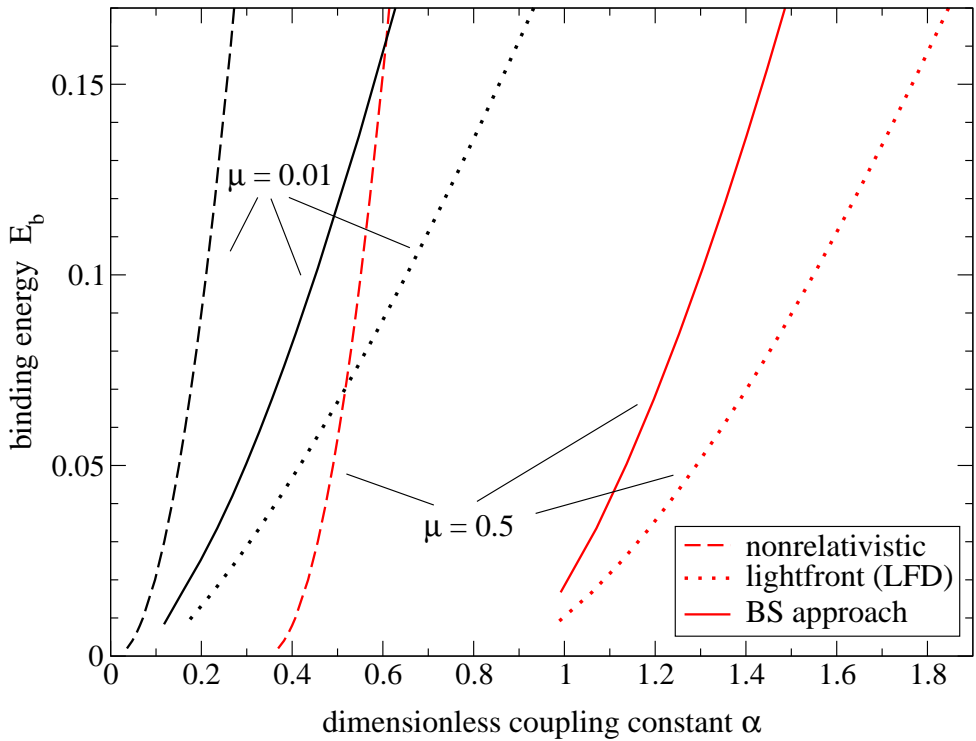}
\caption{(Color online) Three-body binding energy in the limit
when it tends to zero, for $\mu=0.01$ and $\mu=0.5$ (units set by
the constituent mass: $m = 1$).  Solid curves are the BS result;
dotted curves are the LFD results; dashed curves are the solution
of the non-relativistic Schr\"odinger equation.\label{fig5}}
\end{figure}
For small binding energies, our results for the three-body binding
energies $E_b=3m-M_3$ are shown in Fig.~\ref{fig5} (the units are
set by $m=1$).  For comparison, we also include the
non-relativistic result obtained by solving the non-relativistic
Sch\"odinger equation (solved via corresponding non-relativistic
Faddeev equation in the momentum space with the kernel
Eq.~(\ref{yukawa})).  Clearly, all three calculations give
different results.

However, the relativistic binding energies, calculated using the
BS equation and in LFD, are approaching to each other as the
binding energy goes to zero.  Unfortunately, the numerics become
less accurate in that limit, because both the interaction kernels
and the wave functions we are solving for become more peaked.
Nevertheless, for $\mu=0.5$, within numerical accuracy, they
appear to have the same critical coupling, defined as the coupling
constant $\alpha$ at which the binding energy becomes zero.  In
contrast, the non-relativistic binding energy behaves quite
differently, and vanishes at a much smaller value of $\alpha$.

This behavior is similar to that which was found in the two-body
system \cite{mc_2000}.  In the latter case, the decreasingly small
binding energies, calculated via relativistic (both BS equation
and LFD) and non-relativistic equations, become close to each
other only in the limit $\mu\to 0$.

Indeed, for the value $\mu=0.01$ in Fig.~\ref{fig5}, the relativistic
and non-relativistic results approach each other as all the binding
energies decrease down to $E_b\approx 0.01$, and they seem to
extrapolate to $\alpha \approx 0$ as the point at which the binding
energy becomes zero.  We did not calculate for smaller $E_b$ and $\mu$
because of increasing numerical errors.

More intriguing is the rather large difference we see in this
figure between the results from the BS equation and those from LFD
for nonzero binding energy $E_b$.  The LF equation in ladder
truncation significantly underbinds the three-body system relative
to the ladder BS equation. Though at small binding, both BS and
LFD binding energies tend to each other and to zero, the {\em
relative} difference between them remains significant, and is
roughly a factor of two.  We remind the reader that for two-body
system these binding energies approximately coincide with each
other \cite{mc_2000}.

\begin{figure}[b]
\includegraphics[width=0.96\columnwidth]{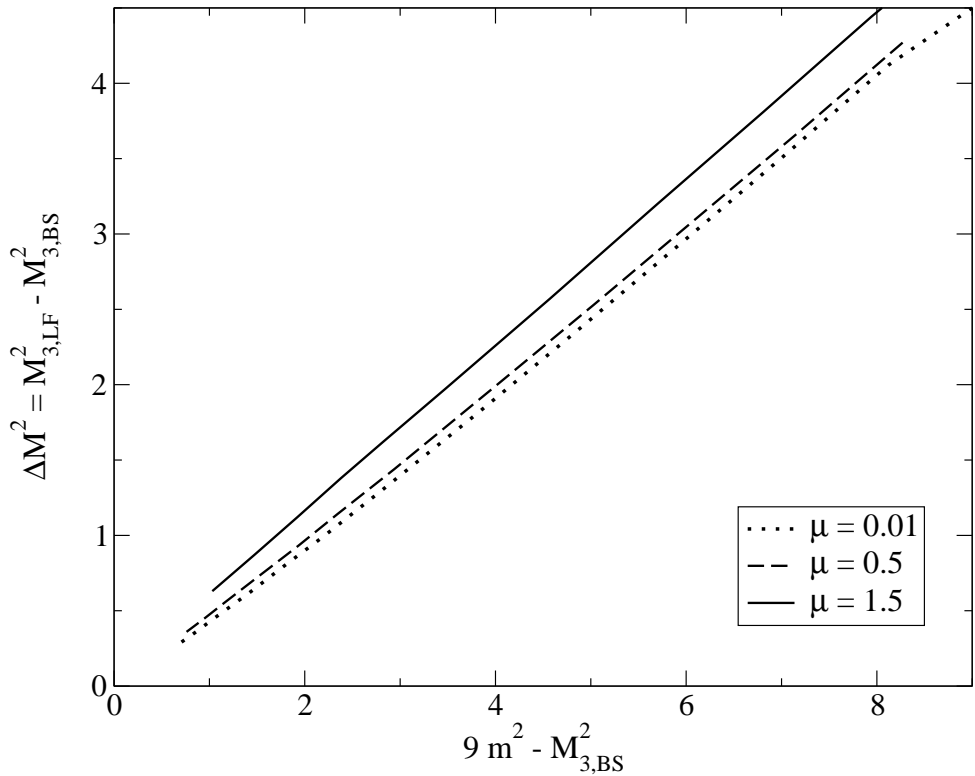}
\caption{(Color online) The difference $M_{3,LF}^2-M_{3,BS}^2$ vs.
the deviation of $M_{3,BS}^2$ from the free value:
$9m^2-M_{3,BS}^2$, for different values of the exchange mass
$\mu$.\label{fig5a}}
\end{figure}

A closer look at our results reveals that the difference between
the three-body masses squared in the BS approach and in LFD,
$M_{3,LF}^2-M_{3,BS}^2$, is approximately proportional to
$M_{3,BS}^2$ itself.  In Fig~\ref{fig5a}, we show the difference
$$
\Delta M^2=M_{3,LF}^2-M_{3,BS}^2
$$
as function of $9m^2 - M_{3,BS}^2$.  It is a remarkably linear
function over the entire accessible interval of $M_{3,BS}^2$
(remember: $9m^2 - M_3^2 = 0$ means zero binding energy, and
$M_3^2=0$ is the maximal binding energy); furthermore, it depends
only weakly on $\mu$, the mass of the exchange boson.

\subsection{Differences largely explained by 3-body force}
\begin{figure}[htbp]
\includegraphics[width=0.96\columnwidth]{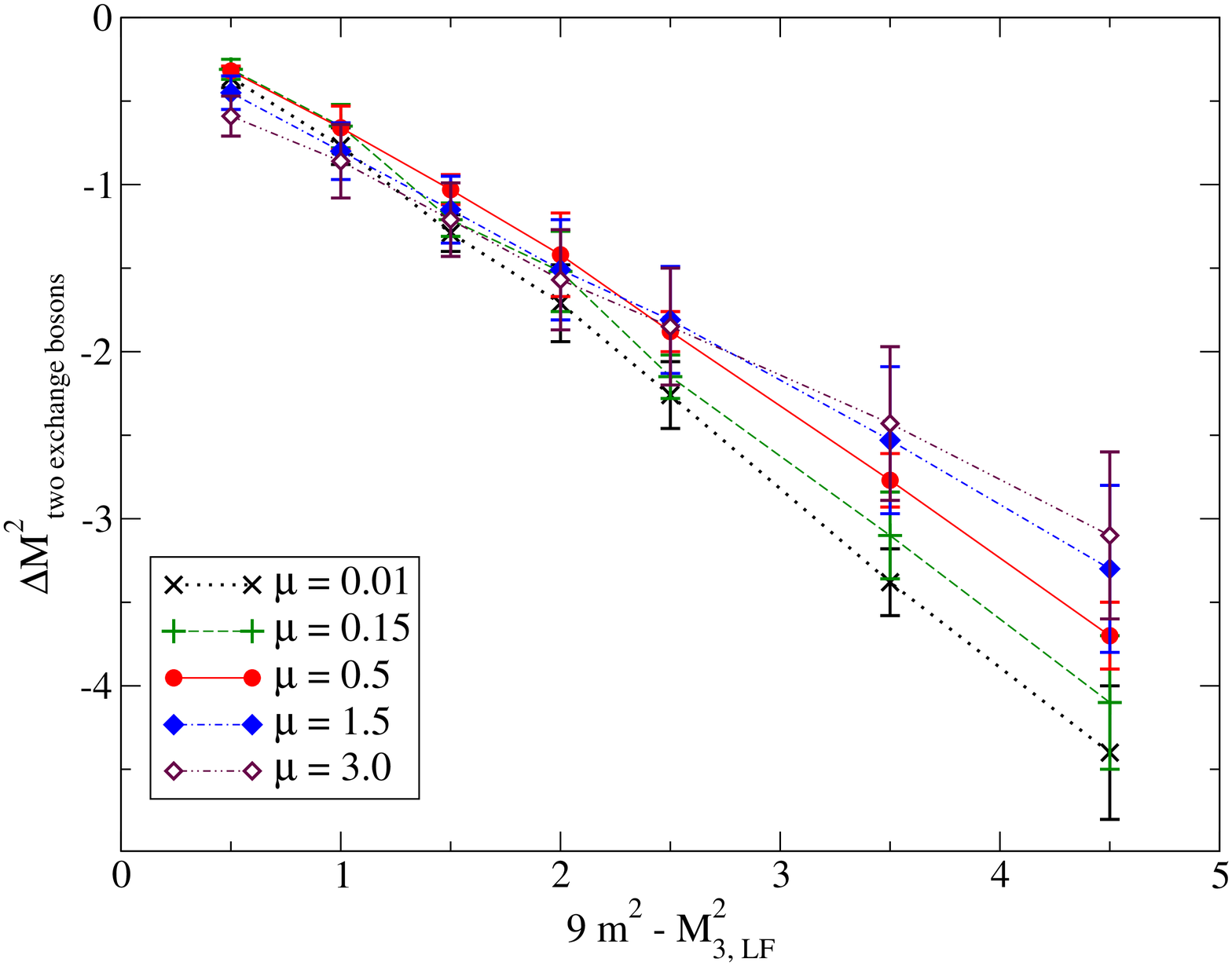}
\includegraphics[width=0.96\columnwidth]{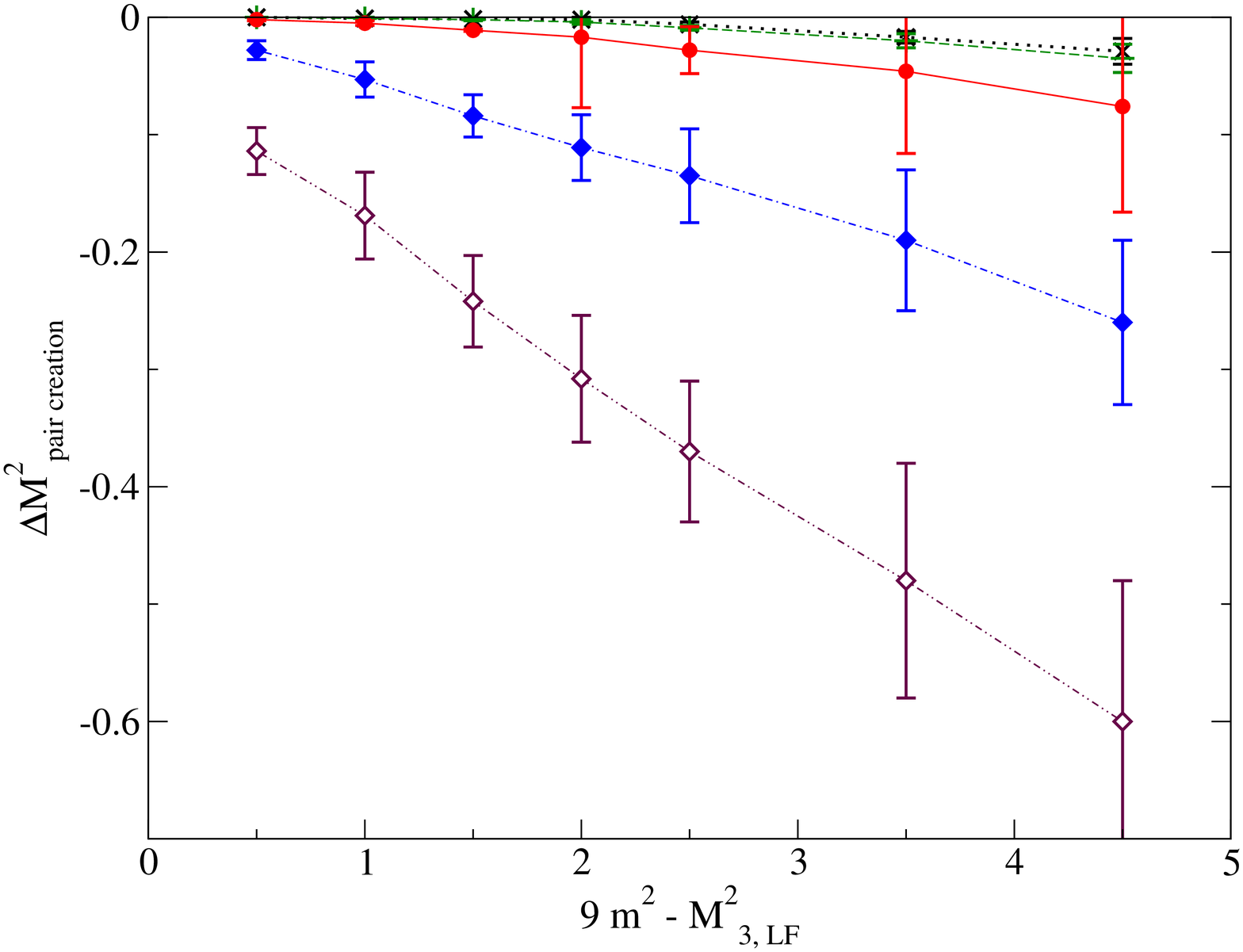}
\caption{(Color online) Correction $\Delta M_3^2$,
Eq.~(\protect{\ref{DM32}}), vs. the difference $9m^2-M^2_{3,LFD}$ for
different values of the exchange mass $\mu$. Top: Contributions from
the first eight graphs in Fig.~\protect{\ref{3bf}} (without
pair-creation); Bottom: Contributions from pair-creation only, the
last diagram in Fig.~\protect{\ref{3bf}}.  Notice the difference
in the vertical scale. \label{fig6}}
\end{figure}

The next question is: Can this difference between the BS approach
and LFD be explained by the diagrams of Fig.~\ref{3bf}?  In order
to answer this question, we performed a perturbative calculation
of the correction to $M_{3,LF}^2$ due to these three-body forces,
using Eq.~(\ref{DM32}).  In Fig.~\ref{fig6} we show our results
for the contributions both from the diagrams with two
exchange-bosons in flight (first eight graphs in Fig.~\ref{3bf})
and from the diagrams with creation and annihilation of an
intermediate particle-antiparticle pair (ninth graph in in
Fig.~\ref{3bf}). Since this calculation involves the numerical
evaluation of an 11-dimensional integral, see Eq.~(\ref{E30}), the
results are not highly accurate; nevertheless, they are quite
illustrative.

As we can see in the top panel of Fig.~\ref{fig6}, the
contribution of two exchange-bosons in flight to $\Delta M_3^2$ is
an almost linear function of $M_3^2$.  Furthermore, it depends only
very weakly on the value of $\mu$ in the wide interval $0.01 \leq
\mu \leq 3.0$.

The contribution of the pair creation term only (ninth graph in
Fig.~\ref{3bf}) is shown in the bottom panel of Fig.~\ref{fig6}.
Again, it is almost linear in $M_3^2$, but, in contrast to two
exchange-bosons in flight, it strongly depends on $\mu$.  At $\mu=3.0$
this contribution is about 20\% relative to two exchange-bosons in
flight but it becomes negligible for $\mu \leq 0.15$.

The behavior of these both contributions is consistent with the
behavior of the difference of the masses squared in the BS and LFD
approaches, Fig.~\ref{fig5a}.  This linearly increasing (in
absolute value) negative correction from effective three-body
forces in LFD eliminates most, if not all, of the linearly
increasing positive difference between BS and LFD results.

\begin{figure}[htbp]
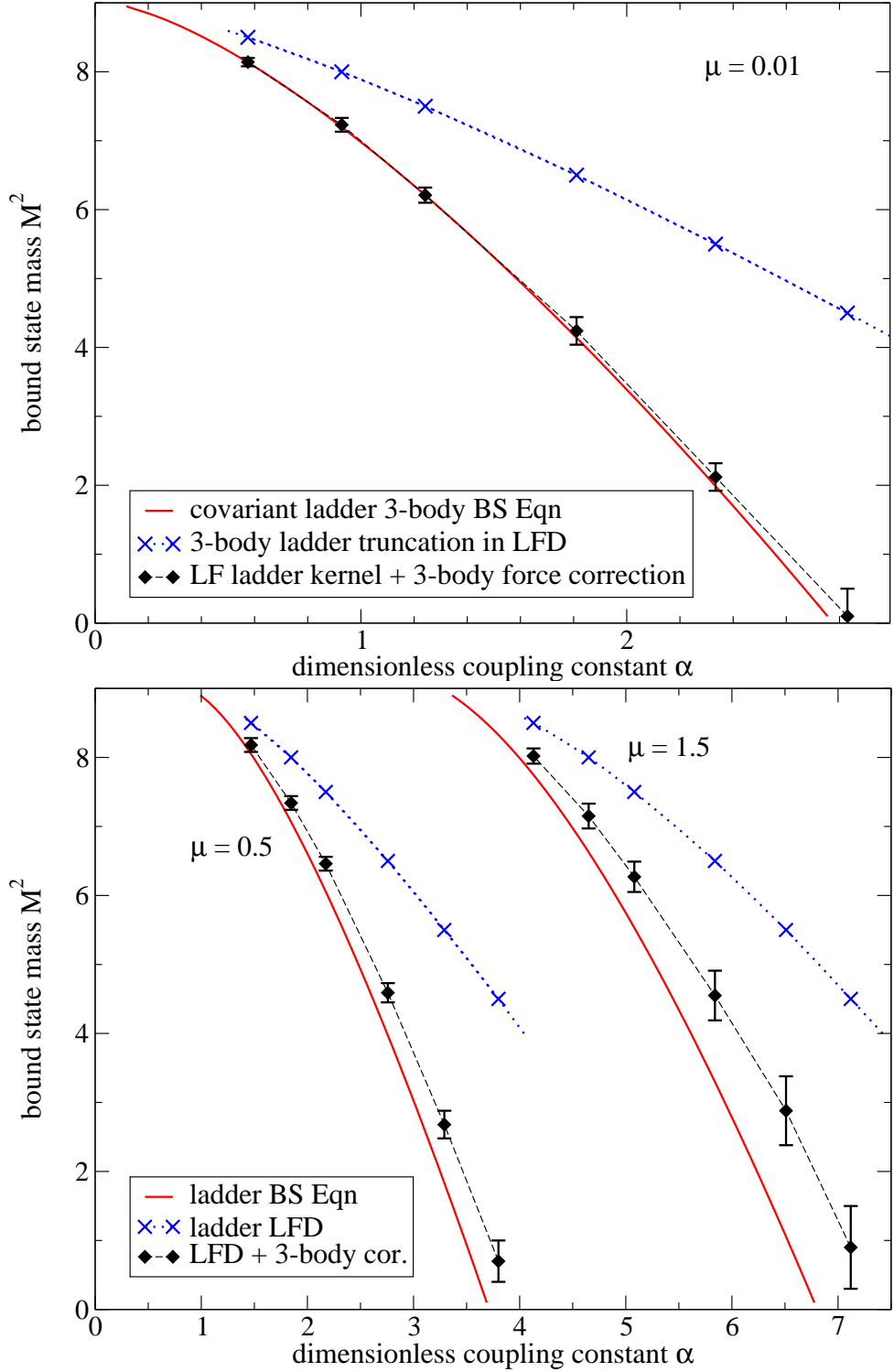

\includegraphics[width=0.96\columnwidth]{resfinal_mu0.01_c.eps}
\includegraphics[width=0.96\columnwidth]{resfinal_mu0.5mu1.5_c.eps}
\caption{(Color online) Three-body bound state mass squared $M_3^2$
vs. coupling constant $\alpha$ for exchange masses $\mu=0.01$ (top),
$0.5$ and $1.5$ (bottom).  The units are set by the constituent mass:
$m=1$.\label{result}}
\end{figure}

The three-body bound state mass squared $M_3^2$ for the exchange
masses $\mu=0.01$, $0.5$ and $1.5$ are shown in Fig.~\ref{result}.
Like in Fig.~\ref{fig5}, the solid and dotted curves are our results
from the three-body BS equation in ladder truncation,
Eq.~(\ref{Eq:bse}), and our results from the LF 3-body bound state
equation with the two-body one-boson exchange kernel,
Eq.~(\ref{kernrx}), respectively.  As in Fig.~\ref{fig5}, the BS and
LFD results differ significantly from each other and the difference
increases with increase of the binding energy (and, hence, with the
coupling constant).

The crosses on the dotted curves indicate the points where we
evaluated the perturbative correction due the 3-body forces depicted
in Fig.~\ref{3bf}.  The diamonds with error bars in Fig.~\ref{result}
indicate the sum $M_{3,LF}^2+\Delta M_3^2$, where $\Delta M_3^2$ is
the correction due from the diagrams of Fig.~\ref{3bf}, calculated
perturbatively by Eq.~(\ref{DM32}), and displayed in Fig.~\ref{fig6}.
The errors indicate the numerical uncertainty, mainly due to the
numerical evaluation of an 11-dimensional integral.  This correction
does indeed shift the LFD three-body mass squared so that it comes in
reasonably good agreement with the BS results.  This is the effect of
relativistic three-body forces, dominated here by two exchange-bosons
in flight, with a small contribution from the creation (and
annihilation) of a pair of constituent particle-antiparticles.

One should of course keep in mind that here we only present a
perturbative estimate of the correction due to these effective
3-body forces in LFD. Note that the perturbative calculation is
unexpectedly good even for large correction compared to the value
of $M^2$ itself, especially for small values $\mu$ of the mass of
the exchange particle. The agreement becomes less pronounced as
the exchange mass increases, once the exchange mass becomes larger
than the constituent mass. E.g. if the exchange mass is three
times the constituent mass, the perturbative correction explains
about 70\% of the difference in $M_3^2$ between the BS approach
and LFD.

Since the corrections are substantial, one could raise the question:
How is a perturbative calculation reliable in this context?  To answer
this question, one should solve equation (\ref{eq3b}) beyond the
perturbative framework, which is numerically much more demanding and
requires significantly more work.  Nevertheless, we have ample
evidence that the contributions of effective 3-body forces in LFD are
substantial in the three-body systems, and can explain most of the
difference between the results found in the BS approach and in LFD.

\section{Conclusion
\label{concl}}
We have solved, for the first time, the three-body BS and LFD
bound state equations with a one-boson exchange kernel and found a
difference in the corresponding binding energies.  This difference
is absent in the two-body BS and LFD equations \cite{mc_2000}.

When the binding energies tend to zero, the absolute difference
between BS and LFD results decreases too.  However, the difference
of the binding energies relative to the binding energy itself
remains roughly constant: LFD in ladder truncation underbinds
approximately by a factor of two, compared to the ladder BS
equation.  The relative difference depends only weakly on the
exchange-boson mass $\mu$.  At the same time, the difference
between the relativistic BS (or LFD) binding energies on one hand,
and the binding energy found from a purely non-relativistic
approach on the other hand, is much larger than the difference
between the BS approach and LFD.  In particular the critical
coupling for the onset of bound states in a three-body system
appears to be the same in the BS approach and in LFD, but in a
non-relativistic calculation this critical coupling is
significantly smaller.  This difference decreases with the
decrease of the exchange-boson mass; in the limit of a massless
exchange boson, this critical coupling tends to zero for all three
approaches.  The latter is quite similar to two-body calculations,
where one also finds a significant difference between a purely
non-relativistic calculation on the one hand, and the BS equation
or LFD on the other hand.

Most of the difference between BS and LFD three-body binding
energies can be attributed to effective 3-body forces. Despite the
fact that both in the BS approach and in LFD we use an one-boson
exchange kernel, the actual kernels in the three-body BS and LFD
bound state equations differ from each other by contributions of
irreducible LFD graphs with two exchange-bosons in flight and by
pair creation.  Both types of graphs are implicitly included in
the ladder kernel of the BS equation (this becomes obvious if its
iterations are transformed in a set of the time-ordered diagrams),
but they are not included in the one-boson exchange kernel in LFD.
In the latter case they should be taken into account separately,
as an extra contribution from a 3-body force.  After taking them
into account perturbatively, the LFD binding energy becomes
approximately equal to the BS results, at least for exchange
masses smaller than the constituent mass ($\mu < m$).  This is a
clear manifestation of three-body forces, which in our model are
unambiguously given by the set of graphs Fig.~\ref{3bf} and
explicitly calculated.

Three-body forces of this origin contribute only to the kernel of
3-dimensional bound state equations, like the LFD one.  They can
also be taken into account as a relativistic correction in the
framework of the Schr\"odinger three-body equation (together with
other relativistic corrections). They are true three-body forces
in a 3-dimensional framework. However, in the framework of the BS
equation, using 4-dimensional integral equations and explicitly
covariant variables without any 3-dimensional reduction, these
effects are automatically incorporated via iterations of the
two-body ladder kernel. They do not manifest themselves as
three-body forces in the BS approach. Therefore, the
interpretation of forces as two- or three-body forces, depends on
the framework with which these forces are associated.

Note that three-body forces, resulting from excitations of
intermediate isobar states, do not depend on the type of equation: BS,
LFD or Schr\"odinger equation.  They remain three-body forces in any
of these equations.  However, they may turn into two-body forces if
one considers the intermediate isobars in the framework of a system of
coupled channel equations, including transitions
$N\leftrightarrow\Delta$.  Both types of these three-body forces could
be taken into account as a correction to a non-relativistic treatment
of nuclei.

In addition, there can be intrinsic three-body forces, such as the
six-nucleon contact term $NNN\to NNN$, generated by chiral
perturbation theory for the effective nucleon interaction at the N2LO
level \cite{N3LO}, or, in QCD, three-body forces due to the triple
gluon vertex.  However, these intrinsic three-body forces are beyond
the skope of the present paper.

We would like to emphasize that although we find a correction to
the LFD kernel (three-body forces) which explains most of the
difference between the LFD binding energy and the BS binding
energy, this does not mean that LFD is considered as an
approximate method to solve the BS equation.  Both the BS approach
and LFD have their advantages and disadvantages.  Without
truncations, they should give the same results for physical
quantities such as the bound state mass.  However, truncations can
lead to differences between LFD and the BS approach. A detailed
comparison between results obtained in LFD and in the BS approach,
such as the work presented here, provides deeper insight in the
limitations and deficiencies of the truncation.  Here, it revealed
that for three-body bound states in the ladder truncation in LFD
there are significant contributions to the binding energy coming
from effective three-body forces related to higher Fock sectors.

Finally, the calculations presented here have been done for spinless
systems: a scalar bound state, composed of three scalar constituents,
interacting via exchange of a scalar boson.  Physical systems such as
the triton and ${}^3$He nuclei, or baryons as bound states of three
quarks, consist of fermions, interacting via (pseudo-)scalar and/or
vector bosons.  We expect that at least qualitatively our conclusions
hold for those systems as well: For three-body systems, the ladder
truncation of the Bethe--Salpeter equation and the ladder truncation
in light-front dynamics are not equivalent, and are likely to give
significantly different results, due to the absence of diagrams like
those in Fig.~\ref{3bf} in the light-front ladder truncation.

\section*{Acknowledgments}
The authors are sincerely grateful to J.~Vary for his interest to this
work, fruitful discussions and support.  One of the authors (V.A.K.)
is indebted for the warm hospitality of the nuclear physics group of
the Iowa State University (Ames, USA), where part of the present work
was performed.  This work was supported in part by the U.S. Department
of Energy Grant DE-FG02-87ER40371.  The work benefited from the
facilities of the NSF Terascale Computing System at the Pittsburgh
Supercomputing Center and from grants of supercomputer time at NERSC.

\appendix
\section{LF Jacobi variables} \label{app1}
The three-body LF wave function depends on the variables
$\vec{k}_{1\perp},\vec{k}_{2\perp},\vec{k}_{3\perp}$ and
$x_1,x_2,x_3$. Because of conservation of momenta, they are not
independent, but satisfy the relations:
$\vec{k}_{1\perp}+\vec{k}_{2\perp}+\vec{k}_{3\perp}=0$,
$x_1+x_2+x_3=1$.

For any sub-system ($\alpha\beta$) we define corresponding LF
Jacobi variables, which are independent from each other.

\subsection{Sub-system 12}
For subsystem 12 we introduce the Jacobi momenta as follows:
\begin{equation}\label{2BV12}
\begin{array}{lcl}
x_{12} &=& \frac{x_1}{x_1+x_2}\ , \cr \vec{k}_{12\perp} &=&
\vec{k}_{1\perp} - x_{12} (\vec{k}_{1\perp}+\vec{k}_{2\perp})\ .
\end{array}
\end{equation}
The construction of $x_{12}$ agrees with general definition of
variable $x$: for subsystem 12 it is the ratio
$x_{12}=\frac{p_{1+}}{p_{12+}}$, where $p_{12+}=p_{1+}+p_{2+}$ is
the plus-component of the four-momentum of the subsystem 12. Dividing
numerator and denominator by the plus-component of the total
three-body momentum $p_{+}\equiv p_0+p_z=p_{1+}+p_{2+}+p_{3+}$, we
find for $x_{12}$ expression (\ref{2BV12}) in terms of the
variables $x_1,x_2$ defined for three-body system.

One can express them as follows
\begin{equation}\label{eq10a12}
\begin{array}{lcl}
x_{12}&=&\frac{x_1}{1-x_3}\ ,
\cr \vec{k}_{12\perp}&=&
\vec{k}_{1\perp}+\frac{x_1}{1-x_3}\vec{k}_{3\perp}
\end{array}
\end{equation}
and vice versa:
\begin{equation}\label{eq10b12}
\parbox{7.5cm}{
\begin{eqnarray*}
x_1=x_{12}(1-x_3),\quad \vec{k}_{1\perp}=\vec{k}_{12\perp}-x_{12}
\vec{k}_{3\perp}.
\end{eqnarray*} }
\end{equation}

The Faddeev component $\psi_{12}$ becomes a function of
$$
\psi_{12}=
\psi_{12}(\vec{k}_{12\perp},x_{12};\vec{k}_{3\perp},x_3). $$ In
contrast to the variables $x_1,x_2,x_3$, constrained by
$x_1+x_2+x_3=1$, both variables $x_{12},x_3$ vary independently in
the interval from 0 to 1.

The three-body integration measure, with integration over two-body
subsystem, is transformed as:
\begin{eqnarray*}
&&\delta^{(2)}(\vec{k'}_{1\perp}+\vec{k'}_{2\perp}+\vec{k}_{3\perp})
\delta(x'_1+x'_2+x_3-1) \frac{d^2k'_{1\perp}dx'_1
d^2k'_{2\perp}dx'_2}{2x'_1 x'_2}
\\
&=&\delta^{(2)}(\vec{k'}_{12\perp}+\vec{k'}_{21\perp})
\delta(x'_{12}+x'_{21}-1)\frac{1}{(1-x_3)}
\frac{d^2k'_{12\perp}dx'_{12} d^2k'_{21\perp} dx'_{21}}{2x'_{12}
x'_{21}}
\\
&\Rightarrow&\frac{1}{(1-x_3)}\,\frac{d^2k'_{12\perp}dx'_{12}}{2x'_{12}(1-x'_{12})}\
.
\end{eqnarray*}

In the non-relativistic limit $x_{12}\approx 1/2$, $x_3\approx
1/3$, and we obtain usual Jacobi coordinates:
\begin{eqnarray}\label{eq10e}
\vec{k}_{12\perp} &\approx&\vec{k}_{1\perp}-\frac{1}{2}
(\vec{k}_{1\perp}+\vec{k}_{2\perp})
=\frac{1}{2}(\vec{k}_{1\perp}-\vec{k}_{2\perp}),
\nonumber\\
\vec{k}_{3\perp} &\approx&\vec{k}_{1\perp}-\frac{1}{3}
(\vec{k}_{1\perp}+\vec{k}_{2\perp}+\vec{k}_{3\perp})
=\frac{2}{3}\left[\vec{k}_{1\perp}-\frac{1}{2}
(\vec{k}_{2\perp}+\vec{k}_{3\perp})\right].
\end{eqnarray}

\subsection{Sub-system 23}
Corresponding formulas are obtained from  the relations  for
the 12-subsystem by the cyclic permutation $123\to 231$. Then the
Jacobi variables are defined as follows:
\begin{equation}\label{2BV23}
\begin{array}{lcl}
x_{23} &=& \frac{x_2}{x_2+x_3}\ ,\cr \vec{k}_{23\perp} &=&
\vec{k}_{2\perp} - x_{23} (\vec{k}_{2\perp}+\vec{k}_{3\perp})\ .
\end{array}
\end{equation}

One can express them as functions of the three-body variables
\begin{equation}\label{eq10a23}
\begin{array}{lcl}
x_{23}&=&\frac{x_2}{1-x_1}\ ,\cr \vec{k}_{23\perp}&=&
\vec{k}_{2\perp}+\frac{x_2}{1-x_1}\vec{k}_{1\perp}
\end{array}
\end{equation}
and vice versa:
\begin{equation}\label{eq10b23}
x_2=x_{23}(1-x_1),\quad \vec{k}_{2\perp}=\vec{k}_{23\perp}-x_{23}
\vec{k}_{1\perp}.
\end{equation}
The Faddeev component $\psi_{23}$ becomes a function of
\[ \psi_{23}= \psi_{23}(\vec{k}_{23\perp},x_{23};\vec{k}_{1\perp},x_1). \]

\subsection{Sub-system 31}
Corresponding formulas are obtained from  the relations  for
the 23-subsystem by the cyclic permutation $123\to 231$ or from the
relations  for the 12-subsystem by the cyclic permutation $123\to
312$. Then the Jacobi variables are defined as follows:
\begin{equation}\label{2BV31}
\begin{array}{lcl}
x_{31} &=& \frac{x_3}{x_3+x_1}\ ,\cr \vec{k}_{31\perp} &=&
\vec{k}_{3\perp} - x_{31} (\vec{k}_{3\perp}+\vec{k}_{1\perp})\ .
\end{array}
\end{equation}
One can also introduce
$x_{13}=1-x_{31},\vec{k}_{13\perp}=-\vec{k}_{31\perp}$.

One can express them as functions of the three-body variables
\begin{equation}\label{eq10a31}
\begin{array}{lcl}
x_{31}&=&\frac{x_3}{1-x_2}\ ,\cr \vec{k}_{31\perp}&=&
\vec{k}_{3\perp}+\frac{x_3}{1-x_2}\vec{k}_{2\perp}
\end{array}
\end{equation}
and vice versa:
\begin{equation}\label{eq10b31}
x_3=x_{31}(1-x_2),\quad \vec{k}_{3\perp}=\vec{k}_{31\perp}-x_{31}
\vec{k}_{2\perp}.
\end{equation}
The Faddeev component $\psi_{31}$ becomes a function of
\[ \psi_{31}= \psi_{31}(\vec{k}_{31\perp},x_{31};\vec{k}_{2\perp},x_2). \]

\subsection{Relation between different Jacobi coordinates and cyclic
permutations}\label{permut} We express the coordinates for the
subsystems 23 and 31 in terms of 12:
$x_{12},\vec{k}_{12\perp},x_3,\vec{k}_{3\perp}$. For that we
substitute in the above formulas the Eqs.~(\ref{eq10b12}) and use the
relations:
\begin{equation}\label{eq10d}
\begin{array}{ll}
x_1=\displaystyle{x_{12}(1-x_3)},&
\vec{k}_{1\perp}=\vec{k}_{12\perp}-x_{12} \vec{k}_{3\perp},
\\
&
\\
x_2=1-x_{12}(1-x_3)-x_3,&
\vec{k}_{2\perp}=-(\vec{k}_{12\perp}-x_{12}
\vec{k}_{3\perp})-\vec{k}_{3\perp}.
\end{array}
\end{equation}
Then we find:
\begin{equation}\label{v23}
\begin{array}{ll}
x_{23}=\displaystyle{\frac{(1-x_{12})(1-x_3)}{1-x_{12}+x_{12}x_3}},&
\vec{k}_{23\perp}=\displaystyle{-\frac{x_3\vec{k}_{12\perp}
+(1-x_{12})\vec{k}_{3\perp}}{1-x_{12}+x_{12}x_3}},
\\
&
\\
x_1=x_{12}(1-x_3),&
\vec{k}_{1\perp}=\vec{k}_{12\perp}-x_{12}\vec{k}_{3\perp}.
\end{array}
\end{equation}
and
\begin{equation}\label{v31}
\begin{array}{ll}
x_{31}=\displaystyle{\frac{x_{3}}{x_{12}+x_3-x_{12}x_3}},&
\vec{k}_{31\perp}=\displaystyle{-\frac{x_3\vec{k}_{12\perp}
-x_{12}\vec{k}_{3\perp}}{x_{12}+x_3-x_{12}x_3}}\ ,
\\
&
\\
x_2=(1-x_{12})(1-x_3),&
\vec{k}_{2\perp}=-\vec{k}_{12\perp}-(1-x_{12})\vec{k}_{3\perp}\ .
\end{array}
\end{equation}

After cyclic permutation $123\to 231$ the variables
$\vec{k}_{12\perp},x_{12}$ turn into $\vec{k}_{23\perp},x_{23}$
defined by Eqs.~(\ref{v23}), and similarly for other variables and
permutations.  In this way one obtains Eq.~(\ref{psitot}) for
total wave function.  Note that the transformations of the
$x$-variables are non-linear, since they are defined through
$x_1,x_2,x_3$ by the non-linear formulas (\ref{2BV12}), (\ref{2BV23}),
(\ref{2BV31}).



\begin{thebibliography}{99}

\bibitem{Maris:2003vk}
Maris, P., Roberts, C.D.:
Int.\ J.\ Mod.\ Phys.\ {\bf E12}, 297 (2003)
[arXiv:nucl-th/0301049].

\bibitem{cdkm} Carbonell, J., Desplanques, B., Karmanov, V.A.,
Mathiot, J.-F.:  Phys. Reports, {\bf 300}, 215 (1998)
[arXiv:nucl-th/9804029]

\bibitem{bpp}
Brodsky, S.,  Pauli, H-C.,  Pinsky, S.:  Phys. Reports, {\bf 301},
299 (1998) [arXiv:hep-ph/9705477]


\bibitem{mc_2000} Mangin-Brinet, M.,  Carbonell, J.: Phys. Lett.,
{\bf B474}, 237 (2000) [arXiv:nucl-th/9912050]

\bibitem{kcm06}
Karmanov, V.A.,  Carbonell, J.,  Mangin-Brinet, M.: Nucl. Phys.
{\bf A790}, 598c, (2007) [arXiv:hep-th/0610158]; Few-Body Systems
(to be published) [arXiv:0712.0971 (hep-ph)]

\bibitem{dorkin}
Dorkin, S.M., Beyer, M., Semikh, S.S., Kaptari, L.P.: Few-Body
Systems {\bf 42}, 1 (2008) [arXiv:0708.2146v1 (nucl-th)]

\bibitem{mck03}
Mangin-Brinet, M.,  Carbonell, J.,   Karmanov, V.A.: Phys. Rev.
{\bf C68}, 055203 (2003) [arXiv:hep-th/0308179]

\bibitem{Pieter}
Maris, P.: \emph{Two and three-body bound states in an explicitly
covariant framework}, presented at the international workshop
LC2006: \emph{Light Cone QCD and nonperturbative hadron physics},
Minneapolis, USA, May 15-19, 2006

\bibitem{tobias}
Frederico, T.:  Phys. Lett. {\bf B282}, 409 (1992)

\bibitem{ck_03}
Carbonell, J.,   Karmanov, V.A.:
Phys.Rev. {\bf C67}, 037001 (2003) [arXiv:nucl-th/0207073]

\bibitem{polyzou}
Lin, T.,  Elster, Ch.,   Polyzou, W.N., Witala, H., Gloeckle, W.:
[arXiv:0801.3210 (nucl-th)];

Lin, T., Elster, Ch.,   Polyzou, W.N., Gloeckle, W.:
Phys. Rev. {\bf C76}, 014010 (2007) [arXiv:nucl-th/0702005];
%

Witala, H.,   Golak, J.,  Skibinski, R., Glockle, W., Polyzou,
W.N.,  Kamada, H.:
Phys. Rev. {\bf C77}, 034004 (2008) [arXiv:0801.0367 (nucl-th)]


\bibitem{sbk}
Schoonderwoerd, N.C.J.,  Bakker, B.L.G.,  Karmanov, V.A.:
Phys. Rev. {\bf C58}, 3093 (1998) [arXiv:nucl-th/9806365]

\bibitem{bs2}
Carbonell, J.,   Karmanov, V.A.:
Eur. Phys. J. {\bf A27}, 11 (2006); [arXiv:hep-th/0505262]

\bibitem{friar}
Friar, J.L.: Nucl. Phys. {\bf A684}, 200 (2001)
[arXiv:nucl-th/9911075]

\bibitem{JISP16}
Shirokov, A.M., Vary, J.P., Mazur, A.I.,  Weber, T.A.: Phys.
Letts. {\bf B644}, 33 (2007) [arXiv:nucl-th/0512105]

\bibitem{brueckner}
Brueckner, K.A., Levinson, C.A., Mahmoud, H.M.: Phys. Rev. {\bf
95}, 217 (1954)

\bibitem{yang}
Shin-Nan~Yang; Phys. Rev. {\bf C10}, 2067 (1974)

\bibitem{glockle}
Shin-Nan~Yang, Gl\"ockle, W.: Phys. Rev. {\bf C33}, 1774 (1986)

\bibitem{WickCutkosky}


Wick, G.C.: Phys. Rev. {\bf 96}, 1124 (1954)\\
Cutkosky, R.E.: Phys. Rev. {\bf 96}, 1135 (1954)

\bibitem{sw} Weinberg, S.: Phys. Rev. {\bf 150}, 1313 (1966)

\bibitem{N3LO}
Entem, D.R.,  Machleidt, R.:
Phys. Rev. {\bf C68}, 041001 (2003) [arXiv:nucl-th/0304018]

\end{thebibliography}
\end{document}